\documentclass[aps,prl,twocolumn,showpacs,preprintnumbers,amsmath,amssymb,superscriptaddress]{revtex4-1}
\usepackage[dvips]{graphicx}% Include figure files
\usepackage{dcolumn}% Align table columns on decimal point
\usepackage{bm}% bold math
\usepackage{color}
\usepackage{amsmath}
\usepackage{amssymb}
\usepackage{amsbsy}
\usepackage[utf8]{inputenc}

\newcommand{\be}{\begin{equation}}
\newcommand{\ee}{\end{equation}}
\newcommand{\bea}{\begin{eqnarray}}
\newcommand{\eea}{\end{eqnarray}}
\def \ket #1{\left| #1 \right\rangle}
\def \bra #1{\left\langle #1 \right \rangle}
\def \scalarprod #1#2{\left \langle #1 \right. \left| #2 \right\rangle}
\def \r {{\mathbf{r}}}
\def \n {{\mathbf{n}}}
\def \mat #1{\underline{\underline{#1}}}
\def \typ {\mathrm{typ}}

\begin{document}
 
\title{Universal Scaling Theory of the Boundary Geometric Tensor in Disordered Metals}

\author{Mikl\'os Antal Werner}
\affiliation{Exotic Quantum Phases ``Momentum'' Research Group,
Department of Theoretical Physics, Budapest University of Technology and Economics, 1111 Budapest, Budafoki út 8, Hungary}

\author{Arne Brataas}
\affiliation{Center for Quantum Spintronics, Department of Physics, Norwegian University of Science and Technology, NO-7491 Trondheim, Norway}

\author{Felix von Oppen}
\affiliation{\mbox{Dahlem Center for Complex Quantum Systems and Fachbereich Physik, Freie Universität Berlin, 14195 Berlin, Germany }}

\author{Gergely Zar\'and}
\affiliation{Exotic Quantum Phases ``Momentum'' Research Group,
Department of Theoretical Physics, Budapest University of Technology and Economics, 1111 Budapest, Budafoki út 8, Hungary}

\date{\today}

\begin{abstract}
We investigate the finite-size scaling of the boundary quantum geometric tensor (QGT) numerically 
close to the Anderson localization transition in the presence of small external magnetic fields. 
The  QGT exhibits universal scaling and  reveals the crossover between the 
orthogonal and unitary critical states in weak random magnetic fields. The flow of the QGT near the critical points determines the critical exponents.
 Critical distributions of the QGT are universal and  exhibit a remarkable isotropy 
 even in a homogeneous magnetic field. We predict  
 universal and isotropic Hall conductance fluctuations at the metal-insulator transition  
 in an external  magnetic field. 
 \end{abstract}


\maketitle
\emph{Introduction.--- }
The geometrical structure of  the Hilbert space continues to receive a lot of attention.
The Fubini-Study metric tensor  of the Hilbert space~\cite{ProvostVallee1980,FischerMetric},  
also referred to as Fisher information metric~\cite{FischerMetric}, provides a natural measure 
of distance in the Hilbert space, related to quantum fidelity  -- a fundamental  concept 
in quantum information science~\cite{QInfo_Review}.
 The Fubini-Study metric tensor~\cite{ZanardiGiordaCozzini2007,Polkovnikov2013} 
has also been used to analyze quantum critical points in many systems including interacting  spin 
models~\cite{ZanardiPaunkovic2006, Garnerone_Zanardi2009}, %{Albuquerque2010, ZhouBarjaktarevic2008}, 
many-body systems~\cite{Buonsante2007,Rigol2013}, %{YouLiGu2007, VenutiZanardi_Hubbard2008,Rigol2013}, 
%impurity models~\cite{WangTroyer2015}, 
and systems  exhibiting topological 
order~\cite{Yang2008,Garnerone_TopQP_2009}. %{Varney2010, AbastoZanardi2008,}. 
Non-adiabatic dynamics 
in  driven quantum systems is also deeply connected  to  the Riemannian metric of the  ground state manifold~\cite{GrandiPolkovnikovSandvik2011}. %{GrandiGritsevPolkovnikov2010,}

Another fundamental geometric concept in quantum theory is the Berry phase~\cite{BerryPhase}, 
a geometric invariant in quantum state manifolds. The presence of a non-trivial Berry curvature leads to interesting quantum interference 
phenomena~\cite{AharonovBohm, Bloch_ZakPhase} %{,Berry_Experiment1}
and impacts the trajectory of wave packets~\cite{Niu1999,Bliokh2008}. The Zak phase~\cite{Bloch_ZakPhase, Zak} -- 
the Berry-phase associated with closed loops in the Brillouin zone -- and its higher dimensional analogues play central  
role in the description of topological insulators~\cite{TopIns_Review, Hafezi, ChangQAHE}. %, , Wang,vKlitzing, Bernevig, }.  
Non-abelian Berry phases of degenerate ground state manifolds, on the other hand~\cite{non-AbelianBP}, can generate spin 
relaxation in the absence of external magnetic fields~\cite{Pablo1}
and underlie realizations of non-abelian statistics in topological superconductors~\cite{Alicea_NatPhys_11}. 
%, and can  be used for geometrical quantum bit operations~\cite{Pablo2}.

The concepts of the Fubini-Study metric tensor and the Berry phase can 
be unified through the so-called quantum geometric tensor (QGT)~\cite{VenutiZanardi2007}. 
Consider a quantum system, whose Hamiltonian $H(\lbrace \phi_i \rbrace)$ and eigenstates 
$\ket{\alpha(\lbrace \phi_i \rbrace)}$
depend smoothly on a set of real  parameters, $\phi_i$. 
%Consequently, the eigenstates   depend also on the parameters $\phi_i$. 
%, and the dependence can be made smooth -- in the absence of 
%eigenstate degeneracy -- by choosing a proper gauge. 
The QGT of the eigenstate 
$\ket{\alpha(\lbrace \phi_i \rbrace)}$ 
at a point $\lbrace \phi_i \rbrace $ in the parameter space  is then defined as
\begin{equation}\label{QGT_def}
 Q^{ij}_\alpha({\bf \phi}) \equiv  \scalarprod{\partial_{\phi_i} \alpha}{\partial_{\phi_j} \alpha} - \scalarprod{\partial_{\phi_i} \alpha}{\alpha} 
  \scalarprod{\alpha}{\partial_{\phi_j} \alpha} \;.
\end{equation}
The matrix $Q^{ij}_\alpha$ is Hermitian and gauge invariant. Its (symmetric) real part  is 
the metric tensor associated with the manifold
$\ket{\alpha(\mathbf{ \phi})}$~\cite{ProvostVallee1980,VenutiZanardi2007}, 
while its imaginary (antisymmetric) part is the Berry curvature 
form~\cite{BerryPhase,VenutiZanardi2007}, whose surface integral yields the Berry phase associated 
with closed loops in  parameter space.

In the present work, we demonstrate that the QGT offers  deep  insight into a 
long-standing problem in condensed matter physics, Anderson's disorder-driven 
metal-insulator (MI) transition in  small external magnetic fields~\cite{Anderson1958,KramerMacKinnonReview93}.
In particular, the structure of the QGT reflects the universality class of the 
Anderson transition. Elements of the QGT display universal finite size scaling close to the metal-insulator 
transition,  and  capture the flow between the orthogonal ($B=0$) and unitary ($B\ne0$) universality classes.
At the transition,  the elements of the QGT have universal distributions, characteristic of the underlying 
symmetry of the  transition, but, surprisingly,  independent of the direction of the external field.  
We predict that these universal fluctuations show up as universal  and isotropic  Hall conductance fluctuations at the metal-insulator transition.

\emph{Mathematical model.--- }
To investigate the effect of small magnetic fields on the properties of eigenstates close to the 
MI transition, we study  disordered non-interacting spinless fermions  on a three-dimensional cubic 
lattice in  external magnetic fields, described by the Hamiltonian
\begin{equation}
\label{Hamiltonian}
  \hat{H} = \sum_{\r} V_{\r} c_{\r}^\dag c_\r - \sum_{\langle \r, \r' \rangle} \left( t_{\r \r'} c_\r^\dag c_{\r'} + H.c. \right) \; . 
\end{equation}
Here $c_{\r}^\dag$ ($c_{\r}$) creates (annihilates) a fermion on  site $\r = (x,y,z)$, and 
 the   $V_\r$ denote independent random variables  uniformly distributed in the interval $[-W/2, W/2]$. 
 The second term in Eq.~\eqref{Hamiltonian} accounts for nearest neighbor hopping, with  
 the  magnetic field incorporated  through the Peierls substitution, $t_{\r \r'} = t \,e^{i 2 \pi A_{\r \r'}}$. 
In homogeneous  fields we use  the gauge of Ref.~\onlinecite{Werner2015}  for the 
bond vector potentials $A_{\r \r'}$, while  in  random  fields the  $A_{\r \r'}$ 
denote independent, uniformly distributed random variables from the interval $[-W_{B}/2, W_{B}/2]$.

Single particle eigenstates of $\hat H$ are usually classified as 'extended' or 'localized'. 
The latter   emerge close to the band edge, and  are separated from the former  at small disorder  by 
so-called mobility edges~\cite{KramerMacKinnonReview93}. Correspondingly, the system is  insulating or metallic 
if states at the Fermi energy are localized or extended, respectively. 

\emph{Universality classes and criticality.--- }
The spatial structure  of localized (extended) states is characterized by the localization  (coherence) length,   $\xi$.  These latter length scales   depend on the energy of the states,  and   diverge at the mobility edge $E_c$
following a power law,  $\xi\sim |E-E_c|^{-\nu}$. 
This divergent behavior  is the ultimate  basis of  single parameter scaling theory~\cite{Abrahams1979}:
assuming that close to the MI transition $\xi$ is the only relevant length scale, 
the zero temperature dimensionless conductance $g=G/(e^2/h) $ of a system  of size $L$  must be  a function of $L/\xi$ only, 
and therefore must obey the scaling  equation, $\partial g / \partial \ln L = \beta(g)$, with $\beta(g)$ a universal 
function.  This beta function has indeed been determined both perturbatively and numerically in the absence of external magnetic field. 
Its universal properties have been convincingly demonstrated~\cite{Abrahams1979,KramerMacKinnon,SlevinOhtsuki2014}.
In the presence of a sufficiently strong time reversal breaking, however, a clearly distinct, but apparently also universal 
 scaling has been observed~\cite{SlevinOhtsuki1997}. 

The beautiful construction of single parameter scaling must therefore necessarily break down 
in  small magnetic fields. A weak external
 magnetic field  generates a magnetic length scale, $L_B$, which is typically much larger than all microscopic length scales. 
 As a consequence, the dimensionless conductance should also depend on the ratio
 $L_B/\xi$, implying a two-parameter dependence, $g = g(L/\xi, L_B/\xi)$, and invalidating 
  the single parameter scaling theory ~\cite{Larkin}. 
 Fortunately, very close to the transition --- or in very large fields ---  we have $L_B/\xi \to 0$. Therefore
  universal scaling is still recovered at criticality,  but with a modified 'unitary' beta function, $\beta \to\tilde  \beta(g)$~ \cite{PascuBoldiGergely}. 
   So far the intriguing cross-over  between  orthogonal and unitary criticality has not been observed systematically 
 in experiments,   but it  has been investigated to some extent  within the non-linear sigma model approach~\cite{Wegner86,Kotliar92}, 
  where the cross-over in a weak magnetic field has been addressed near the orthogonal critical point 
  in $2 + \varepsilon$ dimensions. Perturbative scaling gives a qualitative picture of the cross-over, but the approximate values of the critical exponents
  are in significant disagreement with numerical results~\cite{SlevinOhtsuki2014}. A precise description of criticality near the unitary fixed point therefore
  appears to be beyond the reach  of this perturbative approach. 
Orthogonal-unitary cross-over has been observed numerically in the critical level spacing statistics in Ref.~\onlinecite{Batsch1996}, but 
the violation of the one parameter scaling theory is not addressed in that work. 
As we show now, the quantum geometric tensor  provides an ideal tool to characterize this cross-over.

\emph{Two-parameter QGT scaling theory.--- }
In the spirit of Thouless~\cite{EdwardsThouless1972}, who related the boundary condition dependence of single particle energies 
to the dimensionless conductance, we shall investigate the boundary condition dependence 
of the   single 
particle eigenstates of  Eq.~(\ref{Hamiltonian}), determined by the eigenvalue equation
\begin{equation}
 \ket{\alpha} = \sum_{\r} \alpha(\r) c_\r^\dag \ket{0}\;, \quad \hat{H} \ket{\alpha} = E_\alpha \ket{\alpha} \; .
\end{equation}
%where $\ket{0}$ denotes the vacuum state, $\alpha(\r)$ is the wave function of the single particle 
%eigenstate, and $E_{\alpha}$ is the corresponding single particle energy.
We prescribe here twisted  boundary conditions,
%\begin{equation}
$
\alpha_{\phi}(\r + L \, \mathbf{n}) = %e^{i (n_x \phi_x + n_y \phi_y + n_z \phi_z)} \alpha_{\phi}(\r) \; ,
e^{i \,\mathbf{n}\cdot {\mathbf \phi}} \alpha_{\phi}(\r) \; ,
$
%\end{equation}
with $\mathbf{n} = \{ n_x, n_y, n_z\}$  a vector of arbitrary integers, and ${\bf \phi} = \{ \phi_x, \phi_y, \phi_z\}$ 
collecting  the boundary twists into a single vector.
For a given system size and disorder realization, we can now 
view each eigenstate  as a manifold, $|\alpha\rangle =  |\alpha(\phi)\rangle$, and define the corresponding 
QGT at zero twist, 
$
%|\alpha\rangle\to 
Q^{ij}_\alpha \equiv Q^{ij}_\alpha({\bf \phi} =0).
$
In the presence of time reversal symmetry, the antisymmetric part of the tensor $Q^{ij}_\alpha$ vanishes. Moreover, the 
sum of $Q^{ij}_\alpha$ over occupied states is the  Hall conductance~\cite{ThoulessHallCond}.
The antisymmetric part  of the QGT is therefore a promising  dimensionless indicator of  time reversal symmetry breaking, while   
the diagonal elements of $Q^{ij}$ are reminiscent of  the Thouless number, and turn out to be  in one to one correspondence with it~\cite{SuppMat}.

These observations  lead us to introducing two real parameters for each eigenstate $\ket{\alpha}$,
\begin{equation}\label{g_h_def}
g(\alpha) \equiv \mathrm{tr} \lbrace Q^{ij}_\alpha \rbrace \, , \;\;\;
h(\alpha) \equiv i \left(Q^{xy}_\alpha - Q^{yx}_\alpha \right) \; .
\end{equation} 
These parameters fluctuate strongly for distinct disorder potentials and eigenstates. We therefore  consider their 
\emph{typical} magnitude,   averaged over a 
large ensemble of samples,  
%\begin{eqnarray}\label{typavg_def}
%\log g_{\typ} &\equiv&  \overline{\log |g(\alpha)|} _{E_\alpha \approx E}  
%g_{\typ}(E,L,W,W_B) &\equiv& \exp \left \lbrace \left.\overline{\log |g(\alpha)|} \right|_{E_\alpha = E} \right \rbrace 
%\; ,\nonumber \\
%h_{\typ}(E,L,W,W_B) &\equiv& \exp \left \lbrace \left.\overline{\log |h(\alpha)|} \right|_{E_\alpha = E} \right \rbrace 
%\log h_{\typ} &\equiv&  \overline{\log |h(\alpha)|} _{E_\alpha \approx E}  
%\; ,
%\end{eqnarray}
\begin{equation}\label{typavg_def}
\ln g_{\typ}  \equiv  \overline{\ln |g(\alpha)|} _{E_\alpha \approx E}  
%g_{\typ}(E,L,W,W_B) &\equiv& \exp \left \lbrace \left.\overline{\log |g(\alpha)|} \right|_{E_\alpha = E} \right \rbrace 
, \;\;\;
%h_{\typ}(E,L,W,W_B) &\equiv& \exp \left \lbrace \left.\overline{\log |h(\alpha)|} \right|_{E_\alpha = E} \right \rbrace 
\ln h_{\typ} \equiv  \overline{\ln |h(\alpha)|} _{E_\alpha \approx E}, 
\end{equation}
that are  functions of energy, system size, disorder strength, and magnetic field.
%The size of the ensemble was determined by requiring  $10^7$ eigenstates for each 
%system size, disorder strength, and magnetic field.
As we demonstrate, these quantities behave as  good scaling parameters, and satisfy 
the universal scaling equations
\begin{equation}\label{scaling}
\frac{\partial g_{\typ}}{\partial \log L} = \beta_{g}(g_\typ, h_\typ) \; , \; \frac{\partial h_{\typ}}{\partial \log L} = \beta_{h}(g_\typ, h_\typ) \;.
\end{equation}
The content of Eq.\,\eqref{scaling} is that the logarithmic  size dependence of the typical values $g_{\typ}$ and $h_{\typ}$ is completely independent  of  
microscopic details such as $W$, $W_B$, or the location of the 
Fermi energy, and is solely determined by  $g_{\typ}$ and $h_{\typ}$.

\begin{figure}
 \includegraphics[width = 0.5 \textwidth]{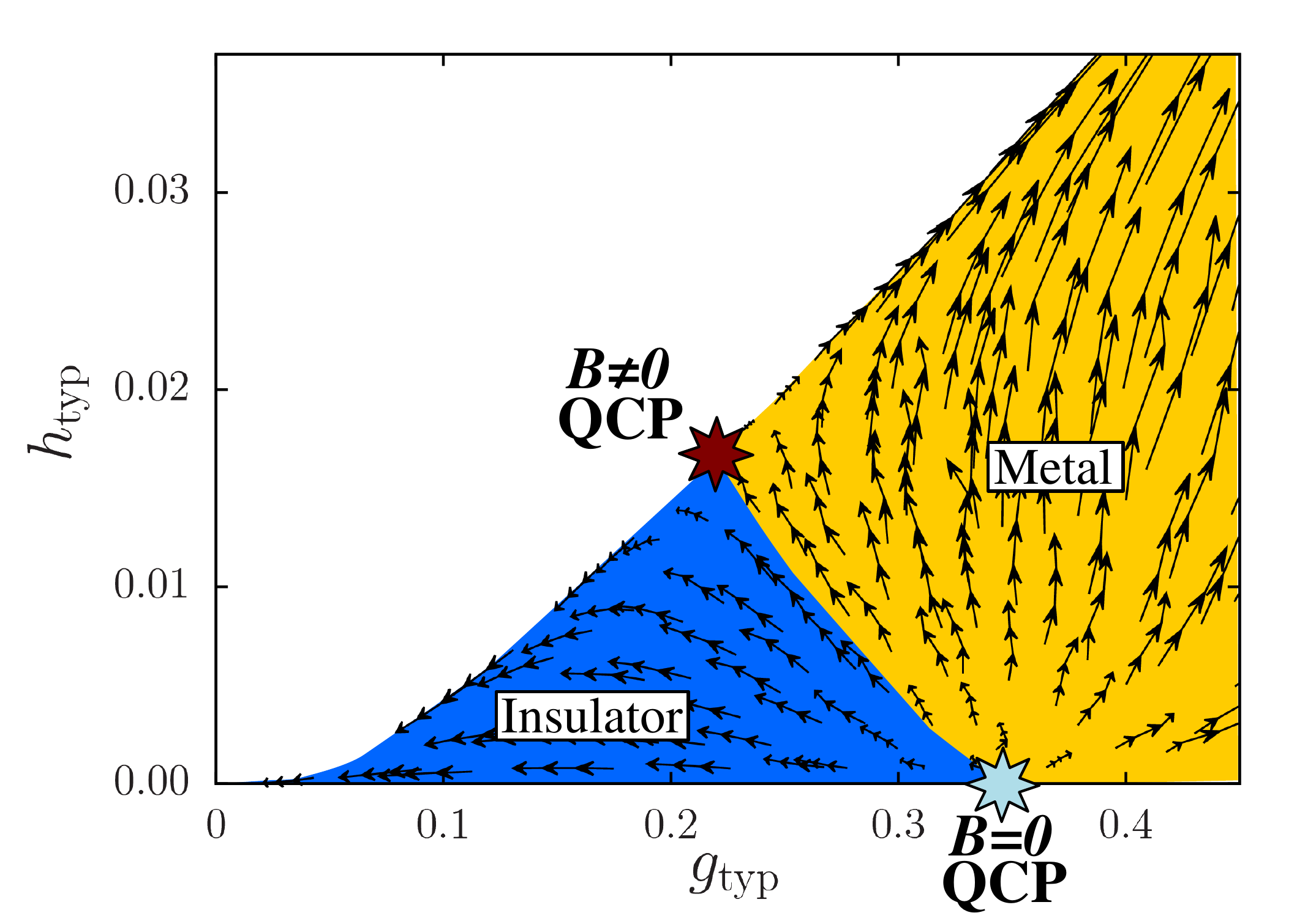}
 \caption{\textit{(Color online)} Finite size scaling of  $g_{\mathrm{typ}}$ and $h_{\mathrm{typ}}$ in a random magnetic field. 
  Arrows indicate the direction of the renormalization group flow upon increasing the   the system size from $L=8$ to $L=14$. The $B=0$ (orthogonal) and  
  $B \ne 0$ (unitary) quantum critical points are denoted by light  and  red stars, respectively. The blue region indicates the insulating phase, 
  while  the metallic phase is yellow.  }\label{fig:flow_randfield}
\end{figure}

To verify the scaling hypothesis of Eq.~\eqref{scaling}, we first performed finite size computations in a homogeneous 
field, and evaluated the logarithmic derivatives on the left hand side for various disorder realizations and energies   numerically.  
In the absence of external field, $h(\alpha) \equiv 0$ for each level, we recover a flow along the axis 
$h_{\typ} =0$ of the $( g_{\typ}, h_{\typ} )$  plane, as shown in Fig.~\ref{fig:flow_randfield}. This flow
is governed by the one-parameter function, $\beta_{g}(g_\typ, 0)$, which we determined numerically
(see Ref.~\cite{SuppMat} for details). 
A critical point emerges at $g_{B=0}^{*} =0.3309(18)$,  and the numerically determined $\beta$-function 
yields a critical exponent $\nu_{B=0} = 1.560(63)$, in good agreement with the best known result for 
orthogonal systems,  $\nu_{\rm orth} = 1.571(8)$~\cite{SlevinOhtsuki2014}. 
We thus find that the \emph{trace} of the QGT, $\sim g$, behaves as 
an appropriate scaling variable, which can be used to replace  the dimensionless conductance 
of the single-parameter scaling theory of Ref.~\onlinecite{Abrahams1979} -- or the Thouless conductance \cite{EdwardsThouless1972}.

\begin{figure*}
 \includegraphics[width = 1.0 \textwidth]{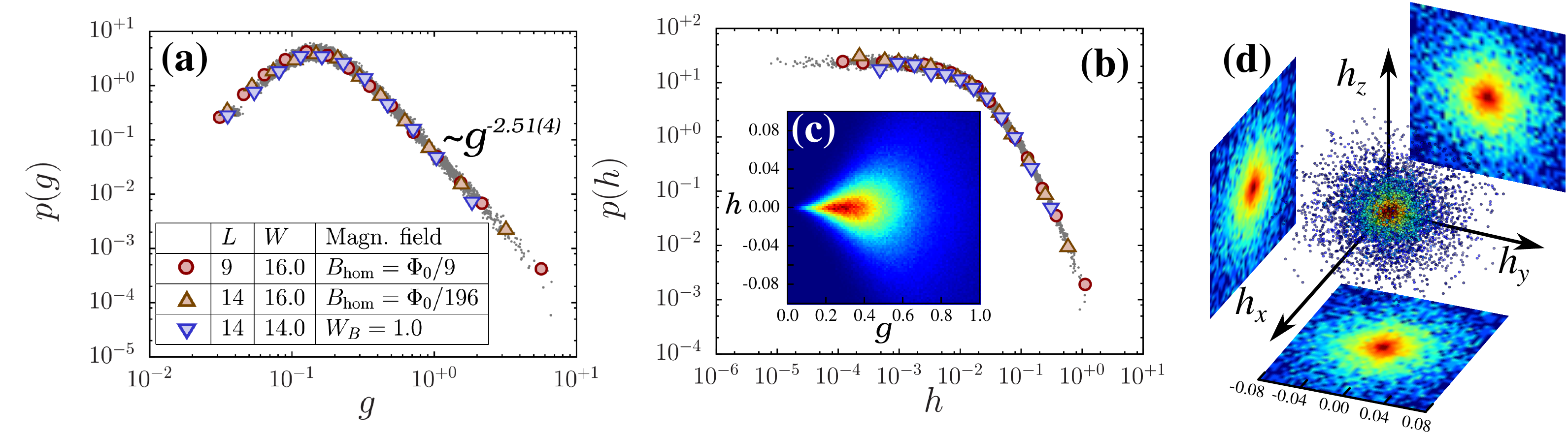}
 \caption{\textit{(Color online)} \textbf{(a)} Probability density function of  $g$ at the unitary ($B \ne 0$) critical point.  Distributions were extracted from systems
 of different system sizes, magnetic fields, and disorder strengths, with the energy at  the mobility edge. 
Different  symbols refer to data extracted from systems 
 with different system sizes, disorder strengths, and homogeneous or random field strengths, as specified in the  legend. 
 The cloud of small gray dots %{\bf (not visible ?????)}
  shows the distribution of $g$, obtained by merging all   data of all 
  parameter sets.
 \textbf{(b)} The $h > 0$ part of the probability density function of the parameter $h$ at the unitary ($B \ne 0$) critical point. Symbols correspond to the 
 parameter sets  as in the panel (a),  while 
 small gray dots show the merged  distribution.
 \textbf{(c)} Joint distribution of  $g$ and $h$ at the unitary  critical point. \textbf{(d)} Joint critical distribution of the  parameters $h_{k}$, 
 characterizing the antisymmetric part of the QGT, determined  at the  unitary fixed point. 
  Points in the  three-dimensional cloud  represent individual eigenstates, computed in a homogeneous magnetic field $B = \Phi_0 / 9$ along the  $z$ direction, 
  in a system of size $L=9$ and disorder $W = 17$. The almost perfect 
 rotational symmetry of the distribution is  supported by the two-dimensional marginals  shown on planes next to the cloud.}\label{fig:prob_g_unit}
\end{figure*}

Piercing just \emph{one} flux quantum through our system brings us immediately to 
the unitary class of systems: It yields  another, universal   one-parameter trajectory in the 
 $( g_{\typ}, h_{\typ} )$  plane, with a critical point at 
$  g_{B \ne 0}^{*} =0.22215(87) $ and a finite  $\;  h_{B \ne 0}^{*} = 0.01683(13)$.  
Again, the extracted value of the critical exponent, $\nu_{B \ne 0} = 1.459(64)$ agrees well with
the most accurate estimate in the literature, $\nu_{\rm unitary} = 1.424(15)$~\cite{UjfalusiLaci}.
These results prove that the antisymmetric part $h$ of the QGT 
 provides a good dimensionless, universal variable to  distinguish the orthogonal and unitary universality classes. 

Unfortunately, within the torus geometry used here, we cannot pierce less than one flux quantum 
through the system~\cite{Werner2015}, and this is already too large to observe
 the flow between the two critical points in the  $( g_{\typ}, h_{\typ} )$ plane. 
 We circumvented this difficulty by applying  \emph{random} vector potentials.  In this case,
 we can tune the strength of time reversal symmetry breaking continuously 
 by changing the  strength $W_B$ of the random vector potentials.
 Reassuringly, in large random fields, $W_B \gtrsim 0.15$, the flow perfectly coincides with the one parameter trajectory that we observed in homogeneous fields. 
 In small random fields ($W_B \lesssim 0.15$), however,  we can now clearly observe a two-parameter 
flow crossing over between   the $B=0$ and $B \ne 0$ universality classes, as presented in Fig.~\ref{fig:flow_randfield}.
We should emphasize that the flow  in Fig.~\ref{fig:flow_randfield} is independent of  microscopic details, 
and is generated at each point by collecting data at different 
energies, and for different values of the disorder parameters $W$ and $W_B$.
A detailed analysis of  the flow around the fixed points   also allows us 
to extract the scaling  exponents associated with the relevant (and leading irrelevant) operator~\cite{Wilson} at the 
orthogonal (unitary) fixed points, associated with time reversal symmetry breaking, 
$$
y_{B=0} = 0.990(11), \;\;\; y_{B \ne 0} = -2.12 (23).  
$$ 

\emph{Critical QGT distributions.--- }
The typical values   $g_{\typ}$ and $h_{\typ}$  still allow for large sample to sample and level to level 
fluctuations   of  $g(\alpha)$ and $h(\alpha)$ at and around the Fermi energy, and a corresponding broad distribution. 
At the critical points  $(g_{\typ}^{*}, h_{\typ}^{*} )$, the Fermi energy lies just on the 
mobility edge, $E_C$, where these distributions are expected 
to become independent of the sample size (scale invariant) and  universal.  
To determine these universal distributions, we first have to locate the mobility edge
for each disorder strength $W$ and $W_B$, and extract the  critical 
distributions of the quantum geometric tensor $Q^{ij}_\alpha$ in its neighborhood. 

 Fig.~\ref{fig:prob_g_unit} summarizes the results for the unitary ($B\ne0$) critical point. 
(For distributions at the $B=0$ orthogonal  critical point see  Fig. S3 of the Supplemental Material~\cite{SuppMat}.) 
The critical distributions of  $g(\alpha)$ and $h(\alpha)$ are indeed  independent of 
system size, disorder, and magnetic field strength. 
The power law tail of the distribution of $g(\alpha)$ resembles that of the critical distribution of the 
Thouless curvature~\cite{Kravtsov_curvature},
though the exponent of the power law is found to be different: while,  in agreement with heuristic arguments presented in the Supplemental Material~\cite{SuppMat},  
$P(g)$ falls off with a power close to $2.5$, the exponent of the Thouless  
curvature's distribution is around $4$~\cite{FelixThoulessPRL, FelixThoulessPRE}. Interestingly,  systems with homogeneous 
and random fields give rise to identical distributions. 
This surprising agreement of the distributions 
in homogeneous and random fields  indicates  that the \emph{direction} of 
the magnetic field is \emph{irrelevant} at the critical point, at least from the point of view of the quantum geometric tensor's structure and distribution.  Therefore 
the statistics of the QGT should be not only universal, but also 
rotationally invariant at the critical point. 

To explore this symmetry, we generalized the 
 parameter $h$ and characterized  the antisymmetric part of $Q^{ij}$ by three independent real numbers 
 forming an axial vector,
\begin{equation}
h_k(\alpha) \equiv i \sum_{i,j} \epsilon_{kij} \, Q^{ij}_\alpha \; ,
\end{equation}
with $\epsilon_{kij}$  the completely antisymmetric tensor. 
%From this point of view, our choice of $h_z$ as the second scaling parameter is rather arbitrary. 
%This step is only \textit{a posteriori} justified: 
As  shown in Fig.~\ref{fig:prob_g_unit}.d, the joint critical distribution of the three parameters $h_{x,y,z}$ shows  remarkable isotropy, even in a strong homogeneous magnetic
field and the typical values  of  $|h_{x,y,z}|$ are all equal. This observation also justifies \emph{a posteriori} the somewhat  arbitrary choice of $h=h_z$  as a scaling variable in a random field, too. 
 
 A detailed analysis of the distribution of the $h_i$, shown in  Fig.~\ref{fig:prob_g_unit}.d -- as well as that of the real symmetrical components of the QGT, 
 shown in the Supplemental Material -- reveals that the distribution of the QGT is not perfectly  O(3) symmetrical, and  slightly breaks rotational symmetry down to a cubic symmetry 
 even at the critical point.
 This small symmetry breaking 
is equally present in random and homogeneous fields, therefore it cannot be induced by the direction of the magnetic field~\cite{SuppMat}, which could anyway only explain the emergence of a tetragonal symmetry. 
Rather, we explain this behavior as an effect of the cubic shape of the system on the structure of 
critical wave functions. 

\emph{Universal Hall conductance fluctuations.--- }
The behavior of the QGT at the critical point has an interesting experimental consequence. 
The antisymmetric part of the QGT is directly related to the  Hall conductance
through the Kubo-Greenwood formula ~\cite{Greenwood,ThoulessHallCond},
\begin{equation}\label{Hall}
 G_{H}^k = \frac{e^2}{\hbar} \sum_{E_\alpha < E_F} h_k(\alpha) \; .
\end{equation}
with $E_F$  the Fermi-energy, and $k \in \lbrace x,y,z \rbrace$  the direction perpendicular to the plane 
of the Hall measurement. According to  Eq.~(\ref{Hall}),  chemical potential changes  in a coherent mesoscopic sample induce 
\emph{Hall conductance  fluctuations}, determined by the  critical distribution of the QGT. 
Consequently, in a magnetic field, at the metal-insulator transition, we predict the emergence of  universal and isotropic  
Hall conductance fluctuations in a mesoscopic sample.  
These fluctuations as well as their precise distributions should be accessible in present-day experiments. As a possible 
implementation, one can think of   disordered  metallic samples in a Hall-measurement setup, with the mobility edge tuned, e.g.,  by applying strain. 
Changing the external magnetic field or the application of back gates should generate the mesoscopic fluctuations discussed here.

%
%In conclusion, we demonstrated that the quantum geometric tensor provides a unified framework to study the Anderson localization transition. 
%The scaling behavior of the QGT confirm the one parameter scaling 
%theory in the orthogonal and unitary universality classes, but in addition to that, it reveals the crossover between them in weak magnetic fields. The critical 
%distribution of the QGT, that is connected to the critical fluctuations of the Hall conductance, 
%show a remarkable isotropy in the unitary critical point.

\emph{Conclusions.--- }
In this work, we demonstrated that the QGT provides a unified framework to capture the 
cross-over between the orthogonal  and unitary  Anderson-localized critical states. Our results show that the geometrical structure of the eigenstates  
is intimately connected to their spatial structure and the conductance properties at the Fermi energy.
 A natural generalization  would be to study the behavior of QGT in models with weak spin-orbit coupling~\cite{UjfalusiLaci,Kawarabayashi}. %{???,}
In that case -- in the presence of time reversal symmetry -- one expects a two parameter 
crossover between the orthogonal and the symplectic classes. If both spin-orbit coupling and magnetic fields are present, an even more complicated, 
three parameter behavior may appear.  It is an intriguing  question if 
the related  cross-overs are reflected in the geometrical structure of the eigenstates, and if the expected
 three parameter scaling can be captured by the QGT. 
Generalizations in the presence of interaction and for  the many 
body localization (MBL) transition are  other open lines of research~\cite{PalHuse, Filippone2016}, though the extremely limited system sizes,  make the scaling analysis 
of  the QGT at the MBL transition a significantly harder task. 

 \emph{Acknowledgement.--- }
We are grateful to Nigel Cooper and Bert Halperin for illuminating discussions. 
This  research  has  been  supported  by the Hungarian National Research, Development and Innovation Office (NKFIH)
 through Grant Nos. SNN118028 and the Hungarian Quantum Technology National Excellence Program (Project No.  2017-1.2.1-NKP-2017- 00001). M.A.W has also been supported by
 the \'UNKP-17-3-IV New National Excellence Program of the Ministry of Human Capacities.

\clearpage

\clearpage
%%%%%%%%%% Prefix a "S" to all equations, figures, tables and reset the counter %%%%%%%%%%
\setcounter{equation}{0}
\setcounter{figure}{0}
\setcounter{table}{0}
\setcounter{page}{1}

\makeatletter
%\captionsetup[figure]{labelfont={bf},labelformat={default},labelsep=period,name={Supplementary Figure}}
\renewcommand{\thefigure}{S\arabic{figure}}
\renewcommand{\theequation}{S.\arabic{equation}}
\renewcommand{\bibnumfmt}[1]{[S#1]}
\renewcommand{\citenumfont}[1]{S#1}
%%%%%%%%%% Prefix a "S" to all equations, figures, tables and reset the counter %%%%%%%%%%

\onecolumngrid

\begin{center}
\Large{\textbf{Supplemental Material to ``Universal Scaling Theory of the Boundary Geometric Tensor in Disordered Metals''}}
\end{center}

\section{Explicit formulas for the boundary geometric tensor}
In this section we {derive} explicit {expressions} for the {quantum geometric tensor (QGT)}~\cite{Polkovnikov}, used in our analysis. The Hamiltonian{,  Eq. (2) of the main text,} 
{is supplemented by} twisted boundary conditions. {In a finite system, this boundary condition appears through the hopping terms at the boundary: there electron operators outside the boundary
($\r'$) are replaced by phase shifted operators inside the boundary, $\r = \r' - \n L$, as
\begin{equation}
 c_{\r'} = e^{i \n \cdot \boldsymbol{\phi}}c_{\r} \; ,
\end{equation}
with $\mathbf{n} = (n_x, n_y, n_z)$ appropriately chosen integers and $\boldsymbol{\phi} = (\phi_x, \phi_y, \phi_z)$ the boundary twists.} {We} can restore the periodic boundary conditions by performing
the gauge transformation,
\begin{equation}
 \tilde{c}_{\r} = e^{-\frac{i}{L} \r \cdot \boldsymbol{\phi}} c_\r \; ,
\end{equation}
 In terms of these, the Hamiltonian becomes
\begin{equation}
 \hat{H} = \sum_{\r} V(\r) \tilde{c}_\r^\dag \tilde{c}_\r - 
 \sum_{\langle \r, \r' \rangle} \left( t_{\r \r'} \;  e^{\frac{i}{L}  (\r-\r') \cdot \boldsymbol{\phi}} \; \tilde{c}_\r^\dag \tilde{c}_{\r'} + h.c. \right) \;.
\end{equation}
To express the QGT, we need the derivatives 
\begin{equation}
 \frac{\partial \hat{H}}{\partial \phi_k} = - \sum_{\langle \r, \r' \rangle} \left( \frac{i}{L}  (r_k-r'_k)  t_{\r \r'} 
 \; e^{\frac{i}{L} (\r-\r') \cdot \boldsymbol{\phi}} \; \tilde{c}_\r^\dag \tilde{c}_{\r'} + h.c. \right) \; .
\end{equation}
{Expanding then $\hat{H}(\boldsymbol{\phi} + d\boldsymbol{\phi}) = \hat{H}(\boldsymbol{\phi}) + d\boldsymbol{\phi} \cdot \frac{\partial \hat{H}}{\partial \boldsymbol{\phi}} + \dots$ and 
performing first order perturbation theory in $d \boldsymbol{\phi}$ we obtain}
\begin{equation}\label{QGT_numerical}
 Q_{ij}(\alpha) = \sum_{\beta \ne \alpha} \frac{\bra{\alpha} \partial \hat{H} / \partial \phi_{i} \ket{\beta} \bra{\beta} \partial \hat{H} / \partial \phi_{j} \ket{\alpha}}
 {\left( E_{\alpha} - E_{\beta} \right)^2} \; .
\end{equation}

\section{The quantum geometric tensor and the Thouless number}
The Thouless number~\cite{EdwardsThouless1972}, {defined as the disorder averaged absolute curvature of the single particle energies at energy $E$, divided by the mean level spacing $\Delta$ at $E$, 
is a commonly used indicator in the field of Anderson transitions,}
\begin{equation}\label{ThoulessNumber}
 {\cal C}_{T} = \frac{\pi}{\Delta}  \left \langle {\left | \frac{\partial^2 E_{\alpha} }{ \partial \phi_x^2} \right |}   \right \rangle _{E_\alpha = E}
\end{equation}
{Similar to} the parameter $g_{\typ}$ (see (6) in the main text), it is a function of energy, system size, disorder strength, and magnetic field.
{As argued} in Ref.~\onlinecite{EdwardsThouless1972}, and numerically demonstrated in Ref.~\onlinecite{BraunMontambaux1997}, 
the Thouless number (\ref{ThoulessNumber}) measures the dimensionless DC conductance of a finite system.

\begin{figure}
 \includegraphics[width = 0.5 \textwidth]{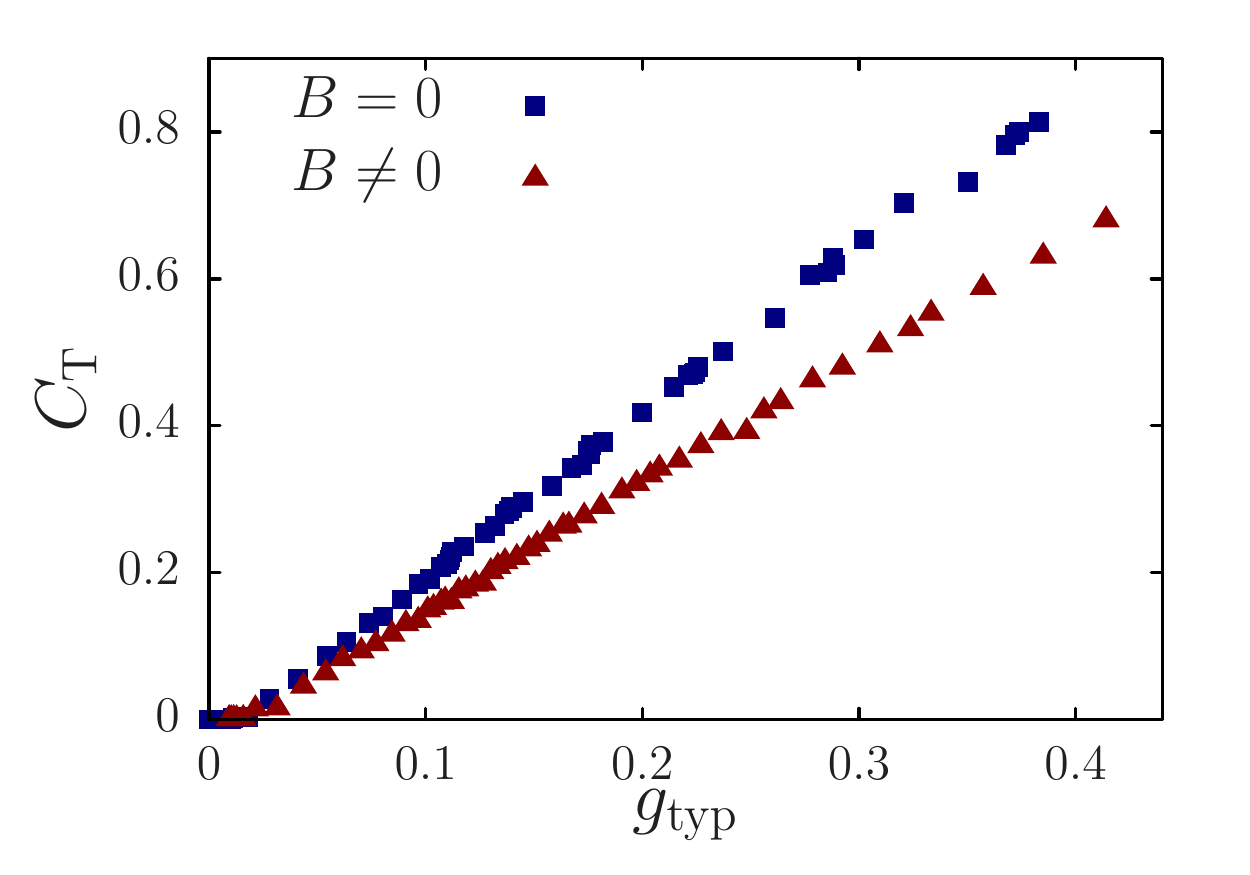}
 \caption{The Thouless number  ${\cal C}_T$ as a function of the parameter $g_{\typ}$: the precise  {relation} between these two parameters 
 {depends} on the universality class. In case of {a} homogeneous or strong random magnetic field, one finds the 
 same  ${\cal C}_T(g_{\typ})$ function, which is clearly different from the one obtained for $B=0$.}\label{fig:g_vs_Thouless}
\end{figure}
Fig.~\ref{fig:g_vs_Thouless} {shows} the connection between the Thouless number and the parameter {$g_{\typ}\sim \mathrm{tr} \, Q$}. We observe an approximately linear connection between the 
two parameters; however, a significant difference {appears} in the {dependence of ${\cal C}_T$ on $g_{\typ}$ in} the presence, or absence of a strong enough external 
magnetic field. Nevertheless, the one-to-one connection between $g_{\typ}$ and  ${\cal C}_T$ in the universal limits implies analogous one parameter scaling properties for both   ${\cal C}_T$ 
and $g_{\typ}$. {Using} $g_{\typ}$ instead of the Thouless number is therefore a legitimate choice.

\section{Finite size scaling of the QGT in a homogeneous magnetic field}
\begin{figure}
 \includegraphics[width = 0.5 \textwidth]{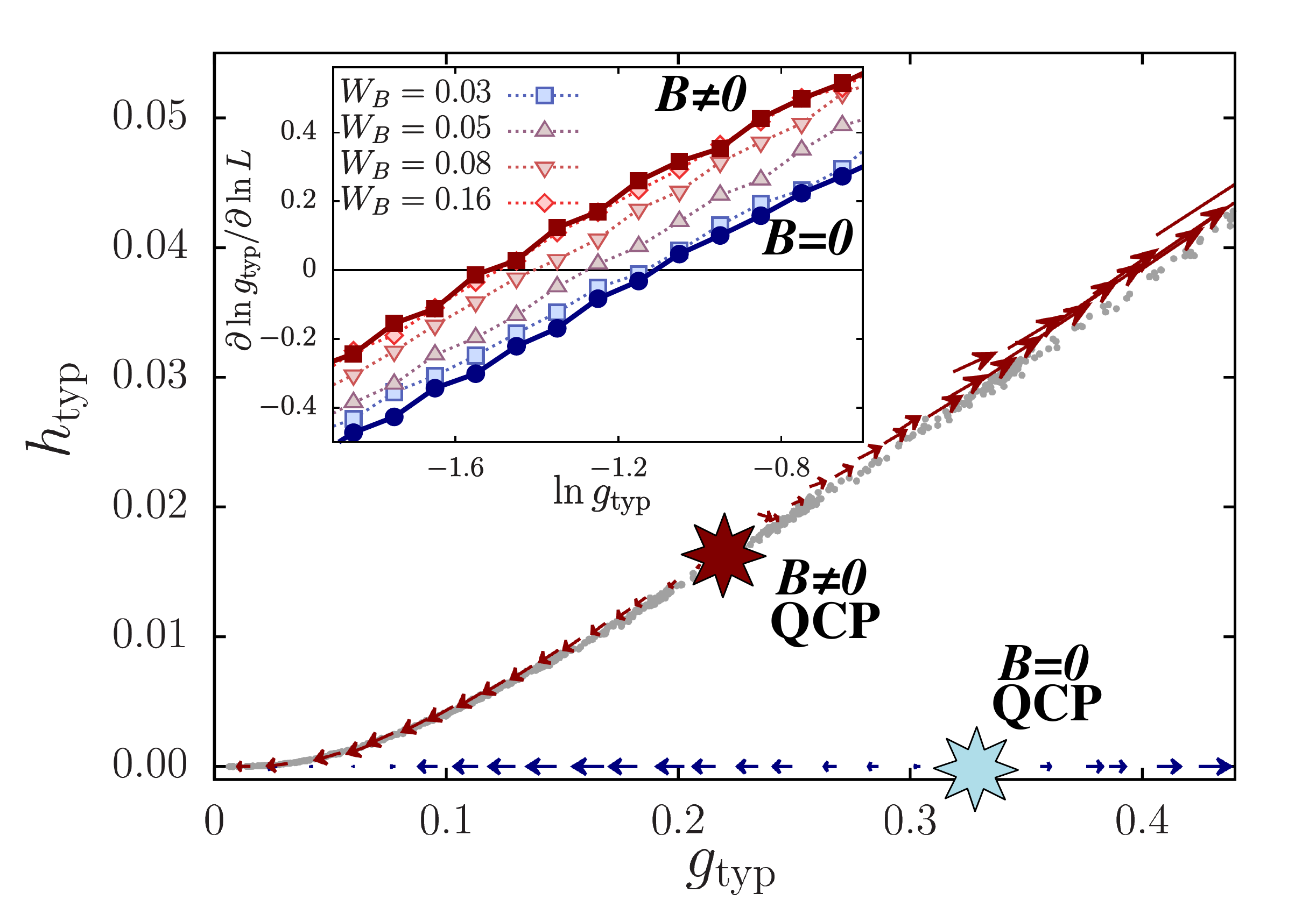}
 \caption{Finite size scaling of {the} parameters $g_{\mathrm{typ}}$ and $h_{\mathrm{typ}}$ in homogeneous (red arrows) and zero (blue arrows) magnetic fields. 
 The renormalization group trajectories fall on distinct one parameter curves in the two cases. 
 The quantum critical points of the $B=0$ (orthogonal) and $B \ne 0$ (unitary) universality classes are denoted by light blue and dark red stars, respectively. 
 The flow in homogeneous field is calculated using a field strength of $B = \Phi_0 / 9$, while the system size is increased from $L=9$ to $L=12$.
 The gray dots show the data in the smallest simulated homogeneous field, $B = \Phi_0 / 196$, with $L=14$. \textit{Inset:} Universal scaling functions 
  $\beta = \frac{\partial \ln g_{\mathrm{typ}}}{\partial \ln L}$ for  $B = 0$ (dark blue circles) and $B \ne 0$ (dark red squares). {Light symbols 
  show the naively calculated $\beta$-functions in weak random magnetic fields, while the system size is increased from $L_1 = 8$ to $L_2 = 14$. These non-universal 
  curves are close to the orthogonal scaling function for the weakest fields but they approach the unitary scaling function as we increase the strength of the random field.}}\label{fig:flow_homorth}
\end{figure}

As stated in the main text, one cannot observe the orthogonal-unitary crossover in homogeneous magnetic fields, because {there is}
at least one flux quantum {pierced} through {a system on a torus}. As shown in Fig.~\ref{fig:flow_homorth}, 
the extracted RG trajectories fall on two distinct lines in the $\lbrace g_{\typ}, h_{\typ} \rbrace$ plane for $B = 0$ and $B \ne 0$.
Even in the smallest magnetic field we could simulate ($B = \Phi_0 / 196$ in the units of flux/cell, and $L = 14$), the 
data points fall on the same line corresponding to the $B \ne 0$ universality class, and we cannot observe any trace of the unitary-orthogonal crossover in 
homogeneous fields. The positions of $B \ne 0$ trajectories and critical points in Fig.~\ref{fig:flow_homorth} coincide with the ones
in Fig. 1 of the main text {computed in a random field}. This agreement strongly supports that models with homogeneous and strong random fields fall in the 
same universality class.

The inset of Fig.~\ref{fig:flow_homorth} shows the one parameter $\beta$-functions ($\beta(g_{\typ}) = \partial \ln g_{\typ} / \partial \ln L$) in the two universality classes.
In addition we show the naively calculated $\beta$-functions in weak random magnetic fields, obtained by calculating numerically the derivative 
$\partial \ln g_{\typ} / \partial \ln L$ by increasing the system size from $L=8$ to $L=14$. We find a ``motion'' of these naive curves from the $B=0$ to 
the $B\ne0$ universal $\beta$ functions {upon increasing the size of the random field.} {This continuous crossover demonstrates the 
{failure of} the one parameter scaling theory in weak random fields: there is necessarily a 
second relevant scaling variable at the orthogonal critical point that describes the crossover.}

\section{Details of fitting the critical points and exponents}
{To} extract the critical parameters of the fixed points in the RG flow {of} Fig. 1. {we} rewrite Eq. (6) into a vectorial 
form and then linearize the equation around the fixed points to get
\begin{equation}\label{eq:lin_flow}
 \frac{\partial}{\partial \ln L} \left( \begin{array}{c} g_{\typ} \\ h_{typ} \end{array} \right) = \left( \begin{array}{cc} M_{gg} & M_{gh} \\ M_{hg} & M_{hh} \end{array} \right) 
 \left[ \left( \begin{array}{c} g_{\typ} \\ h_{typ} \end{array} \right) - \left( \begin{array}{c} g_{\typ}^{*} \\ h_{typ}^{*} \end{array} \right) \right] = 
  \left( \begin{array}{cc} M_{gg} & M_{gh} \\ M_{hg} & M_{hh} \end{array} \right) \left( \begin{array}{c} g_{\typ} \\ h_{typ} \end{array} \right) - 
  \left( \begin{array}{c} b_g \\ b_h \end{array} \right)  \; .
\end{equation}
Here $g_{\typ}^{*}$ and $h_{\typ}^{*}$ denote the coordinates of the corresponding fixed point, while the matrix $\underline{\underline{M}}$ drives the linearized flow. In \eqref{eq:lin_flow} {we 
introduced the}
vector $\left(\begin{array}{c} b_{g} \\ b_{h} \end{array} \right) = \underline{\underline{M}}\left(\begin{array}{c} g_{\typ}^{*} \\ h_{typ}^{*} \end{array} \right) $ to
transform \eqref{eq:lin_flow} {into} a form where the dependence on the parameters $\underline{\underline{M}}$, $b_{g}$, and $b_{h}$ is linear. 
{We can then use} the machinery of multivariate linear regression to extract $\underline{\underline{M}}$ and $\underline{b}$ \cite{MultivariateFit}. 
The coordinates of the fixed point 
are then expressed as $\left(\begin{array}{c} g_{\typ}^{*} \\ h_{\typ}^{*} \end{array} \right) = \underline{\underline{M}}^{-1} \underline{b}$, while the critical exponents of the fixed point are 
the eigenvalues of $\underline{\underline{M}}$.  At the orthogonal ($B = 0$) fixed point, one can {further} exploit the 
fact that in the absence of the magnetic field the parameter $h_{\typ}$ vanishes, {and thereby reduce} the number of fitting parameters from 6 to 4.

\section{The critical distribution of $g(\alpha)$}
\begin{figure}[b]
 \includegraphics[width = 0.8 \textwidth]{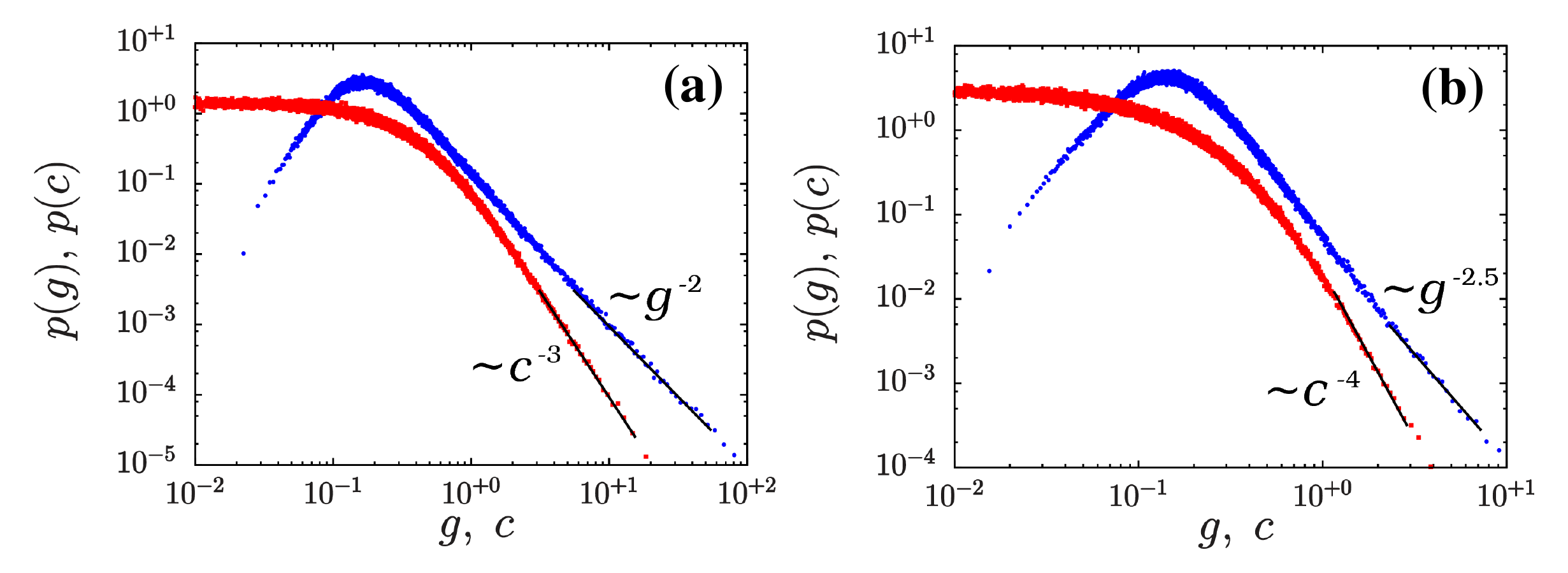}
 \caption{
 \textbf{(a)-(b)}  Critical distributions   of the trace of the geometric tensor,   $g(\alpha) =  \mathrm{tr} \lbrace Q_\alpha \rbrace$,  and the dimensionless level curvature $c_\alpha = \pi \, |\partial^2 E_{\alpha} / \partial \phi^2|/\Delta $  at 
 the orthogonal ($B=0$) and unitary ($B\ne0$) critical points, respectively. Blue  symbols represent 
 the   distribution  $p(g)$,  while red symbols 
represent $p(c)$.   The observed exponents  are consistent with the predictions 
of random-matrix theory.}\label{fig:prob_g_orth_unit}
\end{figure}

We have determined numerically the critical distribution of the trace of the geometric tensor,  $g(\alpha) =  \mathrm{tr} \lbrace Q_\alpha \rbrace$, both at the orthogonal and at the unitary critical points.  The observed $p(g)$ functions, shown in Fig.~\ref{fig:prob_g_orth_unit}, are similar 
at the two critical points, but exhibit important differences, too  (see panels (a) and (b) of
 Fig.~\ref{fig:prob_g_orth_unit}).
At both critical points,  $p(g)$ displays  power-law tails 
$\sim g^{-\eta}$ at large $g$. However,   while  at the unitary critical  point we observe an exponent $\eta\approx 2.5$, 
 at the orthogonal critical point $\eta\approx 2$ seems to emerge. Both exponents are consistent with the 
expression,  $\eta = \beta/2 +3/2$,  with $\beta=1$ and $\beta = 2$  the usual  parameters classifying  the 
orthogonal and unitary universality classes, respectively.  The exponent $\eta$ is clearly different from the exponent 
$\tilde \eta= 2 + \beta$ characterizing the distribution of  the Thouless curvatures, $c_\alpha \equiv\pi\, | \partial_\phi^2E_{\alpha} |/\Delta$, 
defined as the dimensionless level curvature, with   $\Delta $ referring to 
the typical level spacing  (see red data points 
in Fig.~\ref{fig:prob_g_orth_unit}).

The observed exponent  $\eta = \beta/2 +3/2$  follows from the expression Eq.~\eqref{QGT_numerical} under the assumption that  the sum is dominated by a single term with an anomalously small
 level separation,  $s \equiv |E_{\alpha} - E_{\beta=\alpha+1}| \ll \Delta $. By assuming furthermore that small level separations obey Wigner-Dyson statics even at the critical point, 
 $p(s) \propto  s^\beta$, we arrive  immediately at the prediction,   $p(g) \propto g^{-(\beta/2+3/2)}$ for large $g$. 
 Similar arguments imply a fall-off  $p(c) \propto c^{-(\beta +2)}$ for the distribution of the level dependent  Thouless curvature, $c_\alpha$~\cite{curve1,curve1a,curve2}.

\section{Detailed isotropy analysis at the $B \ne 0$ (unitary) fixed point} As stated in the main text, the unitary ($B \ne 0$) critical point has a surprising isotropy: {at} the critical point the direction of a homogeneous magnetic field {appears to be} irrelevant. In addition, the joint distribution of the generalized parameters $h_{k}$ seems to have full rotational symmetry (O(3)) at first sight (see Fig. 2 in the main text).  {Here} we show in more details that, according to our data, the full O(3) symmetry is slightly broken and lowered to the discrete octahedral symmetry, {interpreted as} an effect of the cubic shape of our system.

The QGT, as defined in Eq. (1), follows the usual transformation laws of tensors under rotations: i.e. if one transforms 
the twist phases $\underline{\phi} = (\phi_x,\phi_y,\phi_z)$ with
an orthogonal transformation $\underline{\phi} \rightarrow \mat{O} \; \underline{\phi}$, {then} the QGT transforms as
\begin{equation}
 \mat{Q} \rightarrow \mat{O} \; \mat{Q} \; \mat{O}^T \; .
\end{equation}
{At} the critical points the QGT is a random variable, with a probability density function $P \left( \mat{Q} \right)$. 
The transformation $\mat{O}$ is a symmetry of the distribution, if 
\begin{equation}
 P \left( \mat{Q} \right) \equiv P \left( \mat{O} \; \mat{Q} \; \mat{O}^T\right)
\end{equation}
for every $ \mat{Q}$. 

\begin{figure}
 \includegraphics[width = 0.7 \textwidth]{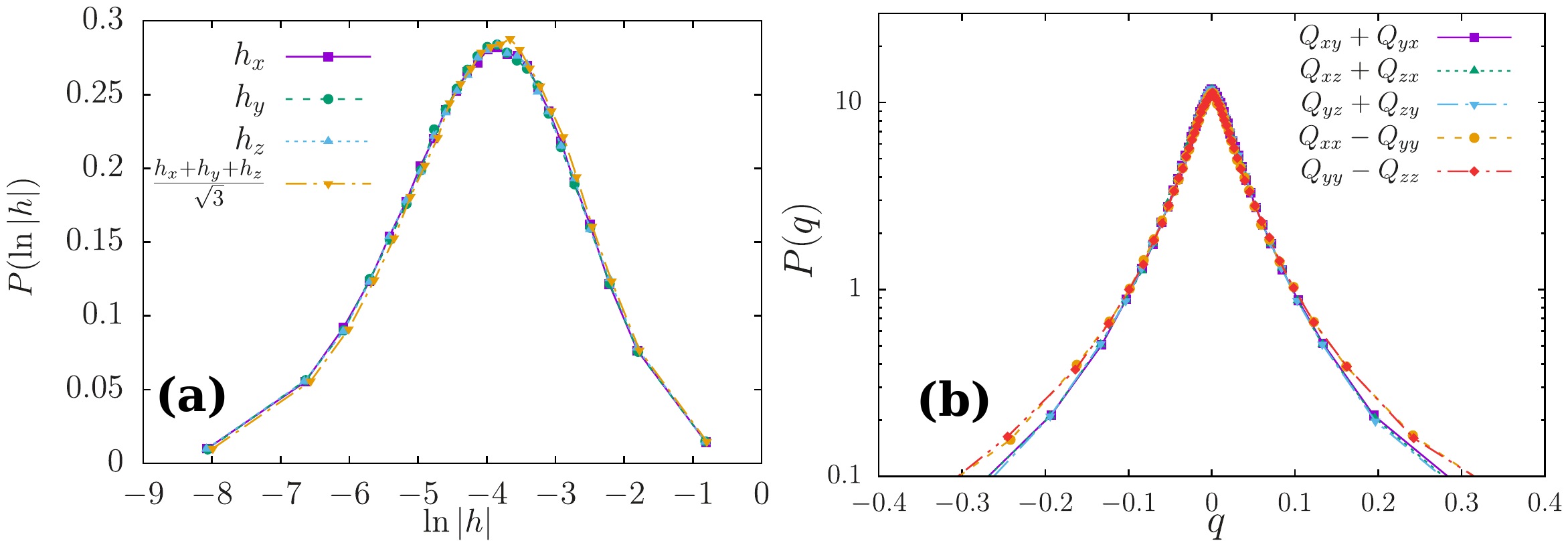}
 \caption{(a) Probability density function of $\ln |h_i|$ and $\ln (|h_x + h_y + h_z|/\sqrt{3})$. In the latter case the distribution slightly distorts, 
 that is an evidence for the breaking of the full rotational symmetry. (b) Probability density functions of 
$Q_{xy} + Q_{yx}$, $Q_{xz} + Q_{zx}$, $Q_{yz} + Q_{zy}$, $Q_{xx} - Q_{yy}$, and $Q_{yy} - Q_{zz}$. The distribution slightly differs for the latter two case, that is again an 
evidence for breaking of the full rotational symmetry. }\label{fig:isotropy}
\end{figure}
The parameters $h_k$ introduced in Eq. (11) parametrize the antisymmetric part of the tensor. {As a result, these parameters follow} the usual transformation laws of (axial-)vectors under 
orthogonal transformations,
\begin{equation}
 \underline{h}' = \det(\mat{O}) \; \mat{O} \, \underline{h} \; .
\end{equation}
Consequently, if the distribution of $\mat{Q}$ had full $O(3)$ symmetry, the joint distribution of parameters $h_{k}$ would be spherically symmetric.
One consequence of such a high symmetry on the marginal distributions would be the relations
\begin{equation}
 P(h_i) \stackrel{?}{=} P(h_j) \stackrel{?}{=} P \left( (h_i + h_j+h_k)/\sqrt{3} \right) \;,
\end{equation}
with $i \ne j \ne k$.
To test this symmetry, {we first} determine the typical magnitudes in the critical point, 
$(h_k)_\typ = 0.01683(13)$, and $\left( (h_i + h_j+h_k)/\sqrt{3} \right)_\typ = 0.01769(11)$. This slight change 
is also directly visible in the probability density function. {The  distribution 
of $\ln |h|$ is shown in Fig.~\ref{fig:isotropy}}, and {only} a slight shift is visible in the case of $(h_i + h_j+h_k)/\sqrt{3}$; however, the distortion of
the distribution is hardly visible.

The breaking of the rotational symmetry is stronger in the symmetric part of the QGT. If the tensor was O(3) symmetric, the following combinations would be 
equivalent,
\begin{equation}
 P \left( Q_{xy} + Q_{yx} \right) , \;  P \left( Q_{yz} + Q_{zy} \right) ,  \;  P \left( Q_{zx} + Q_{xz} \right) ,  \;  P \left( Q_{xx} - Q_{yy} \right) ,  \; \mathrm{and} \;  P \left( Q_{yy} - Q_{zz} \right) \; .
\end{equation}
As shown in panel (b) of Fig.~\ref{fig:isotropy}, for $Q_{xx}-Q_{yy}$ and $Q_{yy}-Q_{zz}$ the distributions are significantly distorted. The separation of the 
diagonal and off-diagonal combinations can be explained as an effect of the lowering of the full O(3) symmetry to the discrete octahedral ($O_h$) symmetry group. 

We believe that in an infinite system the critical point should have full isotropy, i.e. the microscopic direction of the lattice gets irrelevant. In a finite system, however, the critical
state extends over the whole system, and it is therefore unavoidably affected by the boundaries.

\end{document}